\documentclass[twocolumn,showpacs]{revtex4}
\usepackage{times,xspace}
\usepackage{amsbsy,amssymb,amsmath,bm}
\usepackage{graphicx,color,epsfig,rotate}
\usepackage{fancyheadings}

\def\bbbc{{\mathchoice {\setbox0=\hbox{$\displaystyle\rm C$}\hbox{\hbox
to0pt{\kern0.4\wd0\vrule height0.9\ht0\hss}\box0}}
{\setbox0=\hbox{$\textstyle\rm C$}\hbox{\hbox
to0pt{\kern0.4\wd0\vrule height0.9\ht0\hss}\box0}}
{\setbox0=\hbox{$\scriptstyle\rm C$}\hbox{\hbox
to0pt{\kern0.4\wd0\vrule height0.9\ht0\hss}\box0}}
{\setbox0=\hbox{$\scriptscriptstyle\rm C$}\hbox{\hbox
to0pt{\kern0.4\wd0\vrule height0.9\ht0\hss}\box0}}}}

\newcommand{\bi}{{\bf i}}

\newcommand{\ignore}[1]{}
\newcommand{\cComment}[1]{}
\newcommand{\gComment}[1]{}
\renewcommand{\cComment}[1]{\textcolor{blue}{Cristian: #1}}
\renewcommand{\gComment}[1]{\textcolor{red}{Gerardo: #1}}
\begin{document}
\title{Stripes, topological order, and deconfinement in a planar
$t$-$J_z$ model}
\author{J. \v{S}makov$^{1,2,3}$, C.D. Batista$^3$, and G. Ortiz$^3$}
\affiliation{
$^1$Dept. of Physics and Astronomy, Mcmaster University, Hamilton,
Ontario L8S 4M1, Canada\\
$^2$Condensed Matter Theory, Department of Physics,  Royal
Institute of Technology--AlbaNova, SE-10691 Stockholm, Sweden\\ 
$^3$Theoretical Division, Los Alamos National Laboratory, Los Alamos,
NM 87545}
\date{Received \today }

\begin{abstract}
We determine the quantum phase diagram of a two-dimensional bosonic
$t$-$J_{z}$ model as a function of the lattice anisotropy $\gamma$,
using a quantum Monte Carlo loop algorithm. We show analytically that
the low-energy sectors of the bosonic and the fermionic $t$-$J_{z}$
models become equivalent in the limit of small $\gamma$. In this
limit, the ground state represents a static stripe phase
characterized by a non-zero value of a \emph{topological} order
parameter. This phase remains up to intermediate values of $\gamma$,
where there is a quantum phase transition to a phase-segregated state 
or a homogeneous superfluid with dynamic stripe fluctuations
depending on the ratio $J_z/t$.
\end{abstract}


\maketitle

%% Topics to be covered:
%% Historical perspective.
%% Role of topological order vs structure factor.
%% Description and motivation of the model.
%% First time a stripe phase is obtained from a microscopic model 
%% calculation.
%% Methodology: Loop cluster algorithm and largest simulation in the
%% market + adequate method to  calculate the topological order parameter. 
%% Description and characterization of the stripe liquid phase.
%% Comparison with Massimo's results.
%% Comparison with 1d exact results.

During last decade a lot of attention has focused on the study of
inhomogeneous structures in strongly correlated materials, such as the
copper oxide based high-temperature superconductors (HTSC)
\cite{general}. All HTSCs share a common feature -- an
antiferromagnetic (AF) insulating parent state which evolves into a
variety of phases upon doping with carriers, either chemically or using
some external probe. In particular, there is experimental evidence
\cite{expref} that in  part of the phase diagram some of these
compounds feature inhomogeneous charge and spin textures, commonly
known as \emph{stripes}. Existence of such textures is usually
justified by competing interactions between the particle constituents,
which lead to new locally phase-separated states, all of them
characterized by order parameters (OPs) implying that a broken (e.g.,
translational) symmetry state is in place. The real interest lies in
the relation between these stripe phases and superconductivity, since
it is unknown how these quantum orders cooperate with or compete
against each other \cite{nos0}.

%Common wisdom and popularized accounts arguing for the importance of
%those structures justify their existence as a result of competing
%interactions among its particle constituents, such that new locally
%phase separated states of matter emerge which can be distinguished in
%various ways. Some call them stripes, others use more creative names
%to label these potential and appealing phases.

In the present manuscript we argue for the existence of static and
dynamic stripe phases in certain (2+1)-dimensional lattice models of
strongly correlated materials with no charge-ordered state associated
to it, contrary to current understanding. Our starting point is not an
assumed system of interacting {\it stripes} but a system where the
latter emerge from the competition between antiferromagnetism and
delocalization. We demonstrate that the essential physics is related to
the existence of spin antiphase domain structures which are known to be
ubiquitous in various doped AF insulators. The $t$-$J_z$ Hamiltonian is
a minimal model for studying that competition which is also present in
the $t$-$J$ model. Particularly, using numerical simulations we
determine the quantum phase diagram of a hardcore (HC) bosonic
$t$-$J_z$ planar (2D) model, illustrating the main conclusions of this
paper. Our control parameter is the lattice anisotropy $\gamma$,
$0\leq\gamma\leq 1$, with $\gamma=1$ representing the isotropic case
with no explicitly broken lattice rotational symmetry \cite{normand}.
In the small $\gamma$ regime a confinement interaction {\it emerges}
with a non-vanishing topological hidden order and static incommensurate
magnetic orders. This is the static stripe phase which, as the
anisotropy is increased, persists up to a critical value $\gamma_c$
where a confinement-deconfinement transition occurs. At that point the
topological stripe order melts and gives way to a phase-segregated
state for $J_z > J^{c2}_z$ (for $\gamma=1$ see \cite{massimo}) or to  a
superfluid phase with static  AF correlations and dynamical stripe 
fluctuations for $J_z^{c2}> J_z > J^{c1}_z$.

%We address the
%question of whether one can understand the incommensurate magnetic
%response seen in the inelastic neutron scattering experiments without
%such broken symmetry. In particular, we show, using {\it
%exact} numerical simulations, how a stripe liquid phase emerges as a
%function of the anisotropy of the interactions in a hardcore (HC)
%bosonic $t$-$J_z$ model in two spatial dimensions (coupled
%one-dimensional chains).

There is a number of significant reasons to be interested in this
model. First of all, its fermionic counterpart is believed to be a
relevant model for electron motion in the copper oxide planes of HTSCs,
responsible for unique properties of these materials. Second, we
analytically prove that the low-energy spectra of fermionic and bosonic
versions of the model are identical in the limit of weak coupling
$\gamma$ between the chains. This is a natural consequence of an exact
result in one dimension \cite{nos4}. Finally, due to the absence of the
infamous {\it fermion sign problem}, we are able to simulate the
properties of the model for large lattice sizes and very low
temperatures. 

We consider a 2D anisotropic $t$-$J_z$ model on a square lattice for
the spin-$1/2$ HC bosons defined in Ref.~\cite{nos1} 
\begin{eqnarray}
\label{hb1}
H_{B} &=&  \! \sum_{{\bf i},\nu,\sigma} t^{\nu} \left (
{\bar{b}}^\dagger_{{\bf i} \sigma} 
{\bar{b}}^{\;}_{{\bf i}+{\bf e}_{\nu} \sigma}
+ {\rm H.c.} \right ) + H_{J_z} \ ,
\label{hamilt2} \\
H_{J_z}&=&  \sum_{{\bf i}, \nu} J^{\nu}_z
({S}^z_{\bf i} {S}^z_{{\bf i}+{\bf e}_{\nu}} - 
\frac {1}{4}{\bar{n}}_{\bf i} {\bar{n}}_{{\bf i}+{\bf e}_{\nu}}),
\nonumber
\end{eqnarray}
where vectors ${\bf i}=x \ {\bf e}_x+y \ {\bf e}_y$ run over all the
$N_s$ sites of a 2D square lattice, ${\bf e}_\nu$  are the unit vectors
of this lattice, and bars over the operators imply that the constraint
of no more than one particle per site (the HC constraint) has been
already incorporated into the operator algebra. The spin and the number
operators on site ${\bf i}$ are defined by the following expressions
\begin{equation}
\label{ops}
{S}^z_{\bi}=\frac{1}{2}\left(\bar{n}_{\bi\uparrow}-
\bar{n}_{\bi\downarrow}\right)
\,\, \textrm{and} \,\,\,
{\bar{n}}_{\bf i}=\bar{n}_{\bi\uparrow}+\bar{n}_{\bi\downarrow} \, , 
\end{equation}
with $\bar{n}_{\bi\sigma}=\bar{b}^\dagger_{{\bi}\sigma} {\bar
{b}}^{\;}_{{\bi} \sigma}$. Our constrained HC boson operators can be
expressed in terms of standard HC boson operators, ${b}^{\dagger}_{{\bf
i}\sigma}$, ${b}^{\;}_{{\bf j}\sigma'}$, which are HC in each flavor
and obey the commutation relations: $[{b}^{\;}_{{\bf
i}\sigma},{b}^{\;}_{{\bf j}\sigma'}]=[{b}^{\dagger}_{{\bf
i}\sigma},{b}^{\dagger}_{{\bf j}\sigma'}]=0 , [{b}^{\;}_{{\bf
i}\sigma},{b}^{\dagger}_{{\bf j}\sigma'} ]= \delta_{{\bf
ij}}\delta_{\sigma \sigma'} (1-2{n}_{{\bf i}\sigma}) $. The connection
between the two bosonic algebras is given by ${\bar {b}}^\dagger_{{\bf
i} \sigma}= {{b}}^\dagger_{{\bf i} \sigma}(1-{n}_{{\bf
i}\bar{\sigma}})$, where $\bar{\sigma}$ denotes the spin orientation
opposite to $\sigma$. From this expression we can derive the
commutators for the algebra which defines our spin-$1/2$ HC
bosons
\begin{equation}
[\bar{b}^{\;}_{{\bf i}\sigma},\bar{b}^{\dagger}_{{\bf
j}\sigma'} ]=\left\{
\begin{array}{lll}
\delta_{{\bf ij}} (1-2\bar{n}_{{\bf i}\sigma}-
\bar{n}_{{\bf i}\bar{\sigma}}) & \textrm{for} & \sigma=\sigma',\\
-\delta_{{\bf ij}} \bar{b}^{\dagger}_{{\bf i}\sigma'}
\bar{b}^{\;}_{{\bf i}\sigma} & \textrm{for} & \sigma \neq \sigma'.
\end{array}\right.
\end{equation}
Using the general transformations introduced in Refs.~\cite{nos2,nos3},
we can map these 
%spin-$1/2$ 
particles into the HC-constrained fermion
operators, $\bar{c}^\dagger_{{\bf i}\sigma} = {c}^\dagger_{{\bf
i}\sigma} (1-n_{{\bf i}\bar{\sigma}})$ and $\bar{c}^{\;}_{{\bf
i}\sigma} = (1-n_{{\bf i}\bar {\sigma}}) {c}^{\;}_{{\bf i}\sigma}$
\begin{equation}
\bar{c}^\dagger_{{\bf i}\sigma}= \bar{b}^\dagger_{{\bf i}\sigma}
K_{\bf i}^{\dagger}, \;\;\;\;\;\;\;\;\; K_{\bf i}^{\dagger} =
\exp[-i \sum_{\substack{{\bf j}}} \omega({\bf j},{\bf i}) \
\bar{n}_{\bf j}].
\label{trans}
\end{equation}
Here, $\omega({\bf j},{\bf i})$ is the angle between the spatial vector
${\bf j}-{\bf i}$ and a fixed direction on the lattice. Given this
transformation, $H_B$ may be rewritten in terms of the fermionic
operators
\begin{eqnarray}
\label{hb2}
H_{B} &=&  \! \sum_{{\bf i},\nu,\sigma} t^{\nu} \left (
{\bar{c}}^\dagger_{{\bf i} \sigma}e^{iA_{\nu}({\bf i})} 
{\bar{c}}^{\;}_{{\bf i}+{\bf e}_{\nu} \sigma}
+ {\rm H.c.} \right ) + H_{J_z} ,
\end{eqnarray}
with $ A_{\nu}({\bf i})=\sum_{\bf k}[\omega({\bf k},{\bf
i})-\omega({\bf k},{\bf i}+{\bf e}_\nu)] \ \bar{n}_{\bf k}$. 
Expressions (\ref{hb1}) and (\ref{hb2}) for $H_B$ only differ in the
kinetic-energy term, with the fermionic language including a non-local
gauge field $A_{\nu}({\bf i})$ associated with the change in particle
exchange statistics. In general, this gauge field cannot be eliminated
in dimensions larger than one which means that different particle
statistics can give rise to different physics. However, we will see
below that the gauge field is irrelevant and can be eliminated for the
low energy spectrum of $H_B$ in the strongly anisotropic region $|t^y|
\ll |t^x|$. Notice that $t^y$ can be finite and therefore the problem
is still 2D. In other words, when the ratio $\gamma=|t^y|/|t^x|$ is
small the properties of the bosonic model governed by the low-energy
spectrum of $H_B$ will be identical to the corresponding properties of
the fermionic $t$-$J_z$ model [$H_F=H_B(A_{\nu}({\bf i})=\bf 0)$].

\begin{figure}[htb]
\includegraphics[angle=0,width=\hsize,scale=1.0]{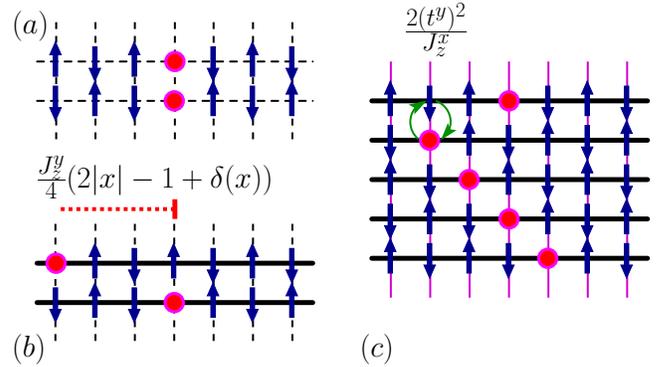}
\caption{
Illustration of the intrinsic stripe formation mechanism in the
2D anisotropic $t$-$J_z$ model (a), and energy costs
associated with different  processes: intra-chain hopping (b) and
inter-chain hopping (c). Dynamical confinement emerges from the
competition between magnetism and delocalization.} 
\label{align}
\end{figure}

Let us start by considering the limit $t^y=0$. In this limit, the
system is an array of one-dimensional (1D) $t$-$J_z$ chains which are
only magnetically coupled through $J^y_z$. From the quasi-exact
solution of the 1D $t$-$J_z$ model \cite{nos4}, it is known that the
lowest energy subspace ${\cal M}_0$ consists of states in which the
spins are antiferromagnetically ordered and each hole is an antiphase
domain for the N\'eel OP (see Fig.~1a). Since the excited subspaces
containing an extra spin excitation are separated by a finite energy
gap, ${\cal M}_0$ becomes the relevant subspace to describe the
low-energy physics of weakly coupled chains. The combination of the
interchain magnetic coupling and the fact that each charge is an
antiphase domain gives rise to a confining interaction between holes on
adjacent chains. This is because the misalignment of holes on the 
neighboring chains either breaks the AF bonds or introduces
ferromagnetic links, both energetically unfavorable
(Fig.~\ref{align}b). The slope of the linear potential is proportional
to $J^y_z=\gamma J^x_z$. This confining interaction leads to a
formation of a hole stripe, which is simultaneously a 2D antiphase
domain boundary for the N\'eel OP \cite{zaanen}. Note that the
characteristic length of stripe fluctuations is  $\ell \sim
\gamma^{-1/3}$. For finite hole densities, $\rho_h$, the stability
condition for the existence of a stripe phase is $\gamma \gtrsim
\rho_h^3 \ t^x/J_z^x$.  

What happens when a small hopping $t^y \ll t^x, J^x_z$ is included? As
shown in Fig.~\ref{align}c, the hopping of the hole to the adjacent
chain has an energy cost of $J^x_z/2$. Since $t^y \ll J^x_z$, the
magnetic structure again acts as a potential barrier confining the hole
to move in the $x$-direction. The main effect of $t^y$ is a second-order 
diagonal correction, $2(t^y)^2/J^x_z$ (see Fig.~1c), to the
energy of the hole. In other words, the low-energy effective model for
$t^y \ll t^x, J^x_z$ is essentially the same as the one for $t^y=0$.
Consequently, the low-energy spectra of the bosonic and the fermionic
$t$-$J_z$ Hamiltonians are the same for $t^y \ll t^x, J^x_z$, i.e., the
gauge field of Eq.~(\ref{hb2}) can be eliminated at low energies
because the hole is effectively moving in the $x$-direction (the
transmutator of statistics $K_{\bf j}^{\dagger}$ is a symmetry of $H_B$
restricted to its low-energy sector). This concludes the proof of the
equivalence of the low-energy spectra of bosonic and fermionic 2D
$t$-$J_z$ models in the strongly anisotropic case. 
%$t^y \ll t^x, J^x_z$.

Since a model defined by the Hamiltonian (\ref{hb1}) is bosonic, its
quantum Monte Carlo (QMC) simulation is not affected by the {\it sign
problem}. We have used the worldline loop algorithm in the continuous
imaginary-time  formulation (for a review see \cite{evertz}), similar
to the one used for the $t$-$J$ model in Ref. \cite{ammon}. This
efficient method allows one to perform calculations at very low
temperatures and large lattice sizes. Unless noted, all simulations
were performed on square lattices of linear size $L$ ranging
from $10$ to $40$, with $\rho_h=0.2$, and error bars smaller
than the size of the symbols. The anisotropy was parameterized by
setting $|t^x|=1$ (all other energy parameters are measured in 
units of $|t^x|$) and introducing a single parameter
$\gamma=|t^y|=J^y_z/J^x_z$, which may vary between zero (disconnected
chains) and unity (isotropic 2D model \cite{massimo}). Temperature was
fixed by putting $\beta\equiv |t^x|/(k_{B}T)=50$, which was low enough
to sample essentially the properties of the ground state.
The method is \emph{exact} up to a statistical error and we are using
\emph{periodic} boundary conditions to avoid the artificial
oscillations introduced by open boundaries \cite{white}.

\begin{figure}[htb]
\includegraphics[angle=0,height=6.0cm,width=9.0cm,scale=1.0]{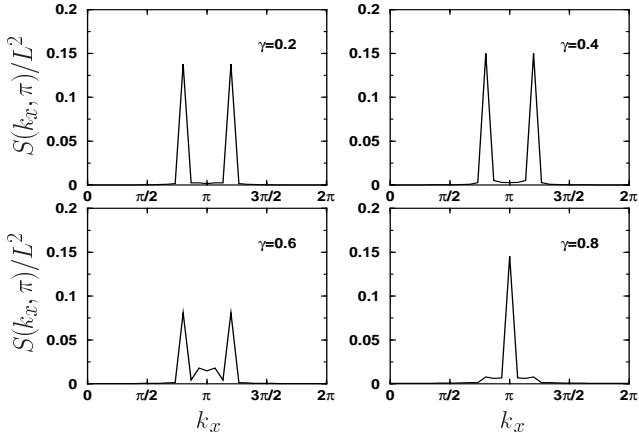}
\vspace*{-0.5cm}
\caption{
\label{s_of_q}
MSF $S({\bf k})$ for a $30\times 30$ system at
hole doping $\rho_h=0.2$, $J^x_z=1$, and different values of anisotropy
$\gamma$. Error bars have been omitted for clarity, standard relative
error for any data point never exceeds 10\%.}
\end{figure}

What observable quantities characterize a stripe phase? As argued
above one needs to measure both the magnetic structure factor (MSF) and a
topological hidden order to uniquely determine it. The MSF ($N_s=L^2$)
\begin{equation}  
S({\bf k})= \frac{4}{N_s} \sum_{{\bf i},{\bf j}} e^{i{\bf k} \cdot
({\bf i}- {\bf j})} \langle {S}^{z}_{\bf i} {S}^{z}_{\bf j}\rangle,
\end{equation}
is expected to display incommensurate peaks at wavevectors ${\bf
k}={\bf Q}\pm(\pi\delta,0)$ in the presence of stripes oriented along
the $y$-direction (across the chains), reflecting the 2D character of
the new antiphase boundaries (see Fig.~\ref{align}c). Here ${\bf
Q}=(\pi,\pi)$ is the AF wavevector. The results for the AF OP are
presented in Fig.~\ref{s_of_q} for $J^x_z=1$. For small values of
$\gamma$, $\gamma=0.2$ and $\gamma=0.4$, there are pronounced peaks at
wavevectors ${\bf k}=(\pi\pm\pi/5,\pi)$ indicating a superstructure
with a period of ten lattice spacings, consistent with incommensurate
spin ordering and a stripe-like phase. At $\gamma=0.6$ the height of
the incommensurate peaks drops while, simultaneously, the peak at ${\bf
k}={\bf Q}$ starts to grow, and becomes dominant for $\gamma=0.8$,
signaling a commensurate AF spin ordering.

The presence of incommensurate peaks in $S({\bf k})$ does not
unequivocally prove the existence of a stripe phase since a similar
behavior can be obtained, for example, for a spin spiral phase. It is
therefore crucial for our analysis to define another quantity which 
unambiguously signals the presence of stripes. Since stripes are
topological defects (antiphase boundaries), this quantity is called
topological OP (TOP) \cite{dennijs} and is calculated for every 1D
chain with fixed $y$ coordinate. The corresponding two-point 
correlation function is
\begin{equation}
\label{top}
G(y)=\frac{4}{N^2_p}\sum^{L-1}_{x,x'=0} \left\langle e^{i\pi
T(x,x',y)} {S}^z_{(x,y)} {S}^z_{(x+x',y)}\right\rangle,
\end{equation}
where ${S}^z_{(x,y)}$ is the spin projection operator at site  ${\bf
i}=x \ {\bf e}_x+y \ {\bf e}_y$, and $N_p=L(1-\rho_h)$ is the number of
particles in the chain. In a system with periodic boundary conditions
Eq.~(\ref{top}) is, of course, independent of $y$, allowing us to
average the results of TOP measurements for different chains. The
parameter $T(x,x',y)$ is defined by
$T(x,x',y)=x'+\sum^{x'}_{p=0}\left(1-\bar{n}_{(x+p,y)}\right)$.
%\begin{equation}
%\label{fact}
%T(x,x',y)=x'+\sum^{x'}_{p=0}\left(1-\bar{n}_{(x+p,y)}\right).
%\end{equation}
Without the second term this parameter would just turn Eq. (\ref{top})
into the square of the 1D N\'eel OP. The second term, however,
introduces an additional factor of $(-1)$ for every hole encountered
between the positions $x$ and $x'$ in the chain with fixed coordinate
$y$, indicating that the hole is an antiphase boundary for the N\'eel
OP. It can be easily established that the TOP defined by (\ref{top})
will reach its maximum value of unity when evaluated in the ground
state of the 1D $t$-$J_z$ model. Thus, $G(y)$ quantitatively measures
to what extent the separate chains in the 2D system have retained
their characteristic 1D ground state topological features. A
non-vanishing TOP {\it and} incommensurate spin ordering in the MSF
constitute strong evidence that the system is in the stripe phase.

\begin{figure}[htb]
\includegraphics[angle=0,height=6.0cm,width=8.8cm,scale=1.0]{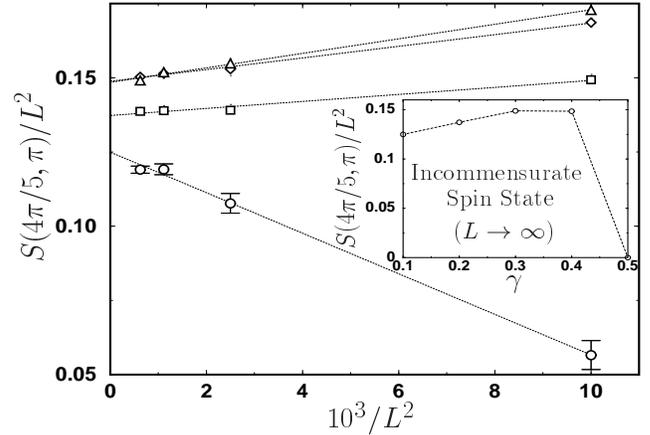}
\caption{
Finite-size scaling of the MSF $S({\bf k})$ at
the position of the incommensurate peak ${\bf k}=(4\pi/5,\pi)$ for
$J^x_z=1$ and different values of the anisotropy $\gamma$: $0.1$
(circles), $0.2$ (squares), $0.3$ (diamonds), $0.4$ (triangles). Lines
are linear fits to the QMC data. Data points in the inset are obtained
by extrapolation to the $L\rightarrow\infty$ limit, with the dashed
line used as a guide to the eye.}
\label{ext_peak}
\end{figure}

\begin{figure}[htb]
\includegraphics[angle=0,height=6.0cm,width=8.8cm,scale=1.0]{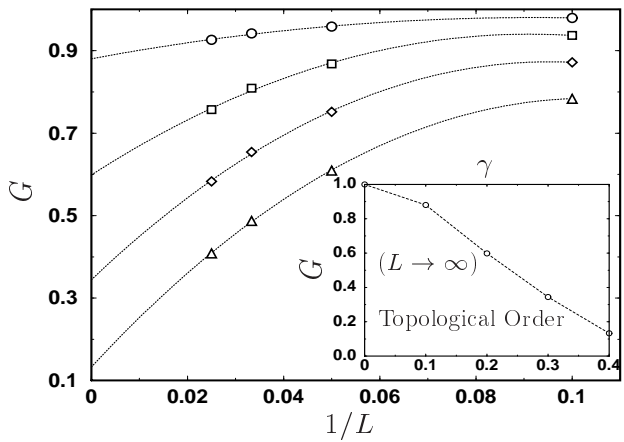}
\caption{
Finite-size scaling of the TOP $G$ for $J^x_z=1$ and different values
of the anisotropy $\gamma$: $0.1$ (circles), $0.2$ (squares), $0.3$
(diamonds) and $0.4$ (triangles). Lines are second-order polynomial
fits to the QMC data. Data points in the inset are obtained by
extrapolation to the $L \rightarrow\infty$ limit, with the dashed line
used as a guide to the eye.}
\label{ext_top}
\end{figure}

In order to rule out the possibility that the observed behavior is a
finite-size effect, we have extrapolated the values of the MSF 
at the position of the incommensurate peak
(Fig.~\ref{ext_peak}) and TOP (Fig.~\ref{ext_top}) for small values of
$\gamma$ to the infinite system size.
Both quantities clearly tend to finite values in the thermodynamic, $L
\rightarrow\infty$, limit for the range of $\gamma$ studied. It is
noteworthy, that  for the same $\gamma$ range the value of the MSF
at the AF wavevector ${\bf Q}$ extrapolates reliably
to zero (within error bars).

The non-uniform behavior of the incommensurate peak height (inset of
Fig.~\ref{ext_peak}) is a consequence of two competing processes. A
stronger coupling between chains leads to a stronger confining
potential, facilitating the formation of stripes. On the other hand,
the same increase tends to disturb the 1D ground state of the chains
due to inter-chain hopping and interactions, driving the system away
from stripe ordering. This competition leads to an optimum value
$\gamma_0\sim 0.3$, for which the peak height reaches its maximum.

\begin{figure}[htb]
\includegraphics[angle=0,width=8.5cm,scale=1.0]{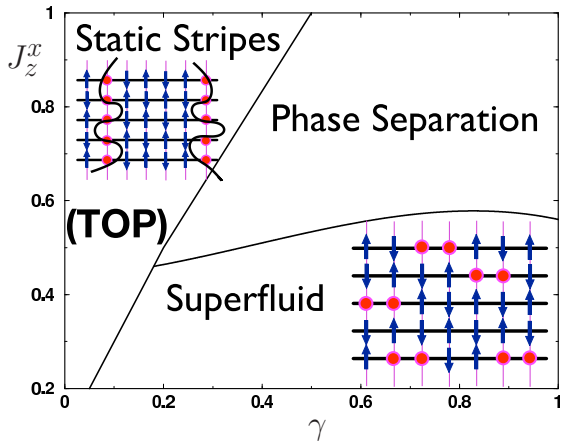}
\caption{Quantum phase diagram of the anisotropic bosonic $t$-$J_z$
model for $\rho_h=0.2$ ($\rho_h^3 \lesssim \gamma \leq 1$), displaying
pictorial representations of the phases.}
\label{quant_ph_diag}
\end{figure}

At $\gamma = \gamma_c$ (see Fig.~\ref{quant_ph_diag}), both the TOP and
the incommensurate peak in $S({\bf k})$ extrapolate to zero in the
thermodynamic limit. For $\gamma>\gamma_c$, we observe nonzero values
of the superfluid density $\rho_s$ with dominant AF correlations if
$J^x_z > J^{c1}_z$ (for $J^x_z < J^{c1}_z$ the system is in the 
Nagaoka ferromagnetic superfluid state). $\rho_s$ was calculated in a 
$10\times10$ cluster by computing the mean square deviation of the
spatial winding number \cite{Ceperley}. Coexistence of AF correlations
and superfluidity indicates that the bosons could be moving in pairs
since the propagation of individual bosons destroys the AF  ordering.
For $J^x_z > J_z^{c2} \sim 0.5$, there is a quantum phase transition to
a phase-segregated state where  a hole-rich superfluid phase coexists
with an AF region with no holes. This is the origin  of the sharp AF
peak of Fig.~\ref{s_of_q} ($\gamma=0.8$). Most likely, the {\it glue}
that keeps particles together in the hole-rich region is still the
confining interaction provided by the magnetic background. 

In summary, we have determined the quantum phase diagram of the planar
anisotropic bosonic $t$-$J_z$ model with periodic boundary conditions. 
For small values of the anisotropy parameter $\gamma$ and hole doping
$\rho_h$, there is a quantum stripe phase characterized by topological
hidden order, and  incommensurate peaks at ${\bf k}= {\bf Q}\pm(\pi
\delta, 0)$ in the MSF. The stripes are formed in
the absence of any physical long-range interactions. The glue
stabilizing each stripe is a confining potential that emerges
dynamically out of the competition between the kinetic energy and the
AF exchange. For anisotropies $\gamma > \gamma_c$  a quantum  phase
transition to a superfluid (deconfining) phase, coexisting with AF
correlations, takes place if $J_z^{c2}> J^x_z > J^{c1}_z$. For larger
values  of $J^x_z$ the system phase segregates. It is remarkable that
the same attractive interaction gives rise to both the stripe phase and
phase segregation.  The existence of a transition between a superfluid
and a phase-separated state in proximity of the stripe instability is
very suggestive when compared with the phenomenology of the HTSC. The
phase-separated region must survive under a transmutation of the
statistics because the fermionic kinetic energy is always higher than
the bosonic one. The presence of an attractive interaction that
segregates the fermions can easily induce a superconducting state if
$J^x_z$ is lower but close to the critical value that leads to the
phase-segregated state. Note that for the bosonic case, this critical
value ($J_z^{c2} \sim 0.5$) is slightly larger than the values of
$J^x_z$ that are considered realistic for the cuprate HTSC.

J.\v{S}. is grateful to the Swedish Foundation for Strategic Research,
Swedish Research Council, G\"oran Gustafsson Foundation, SNAC, and
SHARCNET.
%In the ground state $g(i,j)$ will be equal to one if we have particles at
%both positions $i$ and $j$ and zero otherwise. So, if the chain is in
%the state, in which every hole is an antiphase domain boundary, the TOP will 
%then have it's maximum value of $N_p^2$, where $N_p$ is the number of 
%particles, related to the particle ($\rho_p$) and hole ($\rho_h$) doping by 
%$N_p=N\rho_p=N(1-\rho_h)$.

%It is generally agreed, that in the experiments stripes manifest themselves in
%quasi-two-dimensional systems (such as cuprate oxide high-temperature 
%superconductors) as incommensurate peaks in the magnetic structure factor at 
%${\mathbf k}=(\pi\pm\delta,\pi\pm\delta)$. The observation of these 
%incommensurate peaks in the simulation is not sufficient evidence of
%the stripe phase though, since similar effect may be observed for other microscopic
%spin/charge orderings, in particular, spin density wave (SDW). The TOP allows 
%us to distinguish between these different microscopic states, since non-zero
%value of TOP measured for the coupled chains in the two-dimensional case
%is the evidence of the existence of a confining potential, forcing holes to
%form stripes.

%\begin{figure}
%\epsfxsize=3in
%\epsfbox{fig1.eps}
%\caption{(a) Typical contribution to the ground state of the one-dimensional
%$t$-$J_z$ model. Every hole is the antiphase boundary of an antiferromagnetically
%ordered domain. In a ladder system a state with the aligned holes (b) has lower 
%energy compared to the case with misaligned holes (c)-(e).}
%\label{gs}
%\end{figure}

\end{document}